\documentclass[twocolumn,preprintnumbers,amsmath,amssymb]
{revtex4}
\usepackage{graphicx}
\usepackage[normalem]{ulem}
\usepackage[svgnames]{xcolor}
\usepackage{bm}
\usepackage{multirow}\usepackage{float}
\usepackage{titlesec}

\begin{document}
\title{Universal bound to the amplitude of the vortex Nernst signal in superconductors}

\author{Carl Willem Rischau$^{1}$}
\thanks{Present address: Department of Quantum Matter Physics, University of Geneva, 1205 Geneva, Switzerland}
\author{Yuke Li$^{1}$}
\thanks{Present address: Department of Physics and Hangzhou Key Laboratory of Quantum Matter, Hangzhou Normal University, Hangzhou, 311121, China}
\author{Beno\^{\i}t Fauqu\'e$^{2}$}
\author{Hisashi Inoue$^{3}$}
\thanks{Present address:National Institute of Advanced Industrial Science and Technology (AIST), Tsukuba 305-8565, Japan}
\author{Minu Kim$^{3}$}
\thanks{Present address: Max Planck Institute for Solid State Research, Heisenbergstrasse 1, 70569 Stuttgart, Germany}
\author{Christopher Bell$^{3}$}
\thanks{Present address: H. H. Wills Physics Laboratory, University of Bristol, Tyndall Avenue, Bristol, BS8 1TL, UK}
\author{Harold Y. Hwang$^{3}$}
\author{Aharon Kapitulnik$^{3}$}
\author{Kamran Behnia$^{1}$}
\email{kamran.behnia@espci.fr}
\affiliation{(1) Laboratoire de Physique et d'\'Etude des Mat\'eriaux\\ (ESPCI Paris - CNRS - Sorbonne Universit\'e), PSL Research University, 75005 Paris, France\\
(2) JEIP, USR 3573 CNRS, Coll\`ege de France, PSL Research University,  75005, Paris France\\ 
(3) Geballe Laboratory for Advanced Materials, Stanford University, Stanford, CA 94305, USA\\}

\date{\today}
\begin{abstract}
A liquid of superconducting vortices generates a transverse thermoelectric response. This Nernst signal has a tail deep in the normal state due to superconducting fluctuations. Here, we present a study of the Nernst effect in  two-dimensional hetero-structures of Nb-doped strontium titanate (STO) and in amorphous MoGe. The Nernst signal generated by ephemeral Cooper pairs above the critical temperature has the magnitude expected by theory in STO. On the other hand, the peak amplitude of the vortex Nernst signal below $T_c$ is comparable in both and in numerous other superconductors despite the large distribution of the critical temperature and the critical magnetic fields. In four superconductors belonging to different families, the maximum Nernst signal corresponds to an entropy per vortex per layer of  $\approx$ k$_Bln2$. 
\end{abstract}

\maketitle

Superconducting vortices are quanta of magnetic flux with a normal core surrounded by a whirling flow of Cooper pairs~\cite{Tinkham1996}. In a `vortex liquid' a charge current and an electric field can be simultaneously present and produce dissipation. This state of matter is prominent in high-$T_c$ cuprates~\cite{Blatter1994}. One property of the vortex liquid is a finite Nernst effect (the generation of a transverse electric field by a longitudinal thermal gradient)~\cite{Behnia2016}. Together with its Ettingshausen counterpart (a transverse thermal gradient produced by a longitudinal charge current), it has been widely documented in both conventional~\cite{Huebener1979} and  high-$T_c$ superconductors~\cite{Palstra1990,Li1994,Wang2006}. In the latter case, the debate has been mostly focused on interpreting the persistence of a Nernst signal above the critical temperature~\cite{Wang2006, Behnia2009, Cyr2018}. The vortex origin of the peak signal below $T_c$ remains undisputed and its quantitative amplitude unexplained. Theoretical tradition has linked the magnitude of the finite Nernst signal to the motion of vortices under the influence of a thermal gradient due to the excess entropy of the normal core ~\cite{Stephen1966,Maki1971,Huebener1979,Sergeev_2010}. As a consequence, the magnitude of the Nernst response is expected to strongly vary among different superconductors~\cite{Stephen1966,Maki1971,Sergeev_2010}. 

Here we present a study of the Nernst effect in two superconductors, namely two-dimensional Nb-doped SrTiO$_3$ and $\alpha$-MoGe. We will show that the magnitude of the fluctuating Nernst response above $T_c$ is in agreement with theoretical expectations, but not the amplitude of the vortex Nernst signal in the flux flow regime below the critical temperature. Putting under scrutiny available data for other superconductors (with a range of critical temperatures extending over three orders of magnitude), we find that the observed  peak does not exceed a few $\mu$V/K. Available theories~\cite{Stephen1966,Maki1971,Sergeev_2010} link the amplitude of the vortex Nernst response in a given superconductor to its material-dependent length scales in disagreement with our observation. 

\begin{figure*}
\centering
\includegraphics[width=\linewidth]{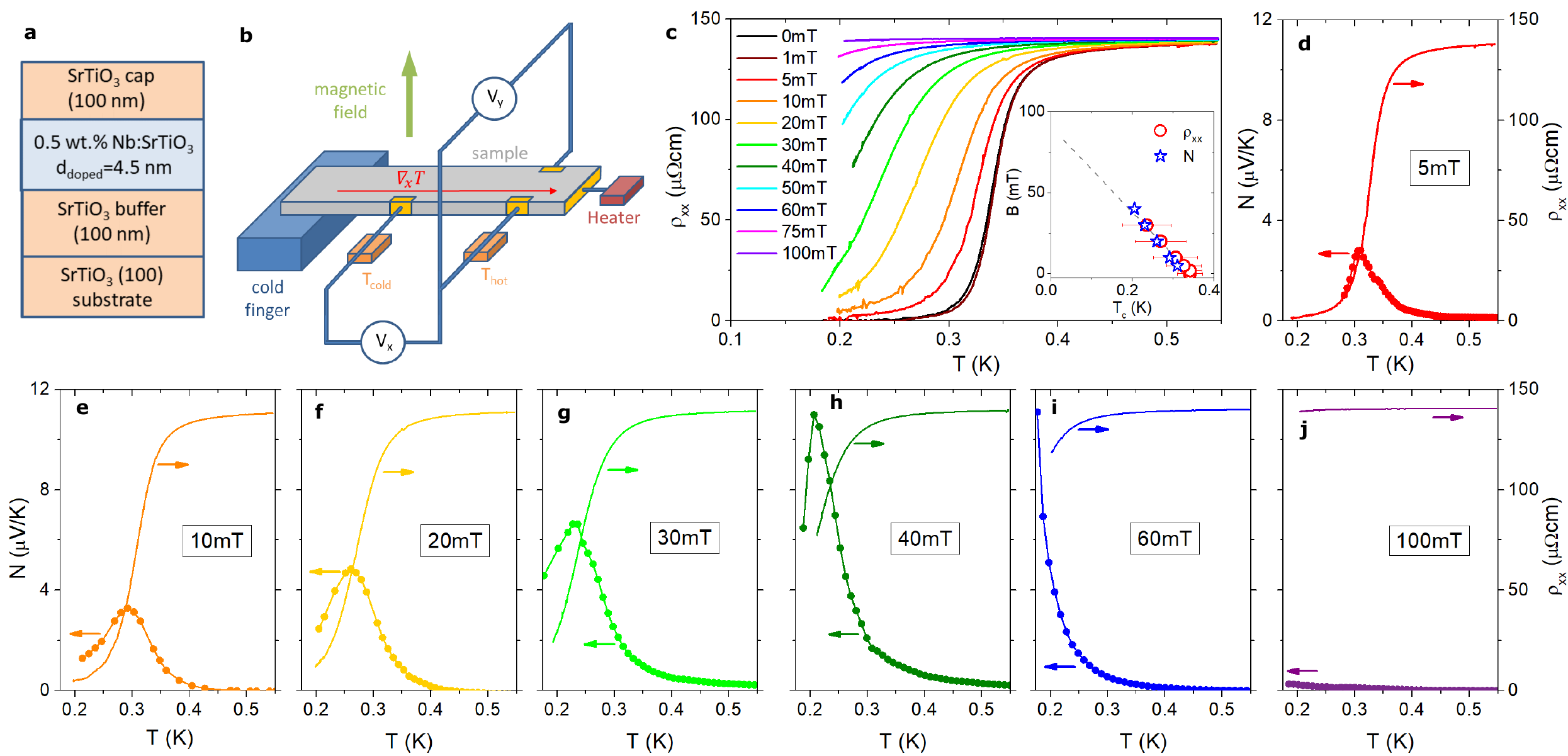}
\caption{\textbf{Nernst effect in two-dimensional Nb-doped strontium titanate:} a) Schematic view of the heterostructure. b) Sketch of the two-thermometers-one-heater set-up used in these measurements. c) Resistivity $\rho_{xx}$ as a function of temperature. The midpoint resistive transition at $T_c=0.341$ K shifts to lower temperatures with increasing magnetic field. The inset shows the correlated evolution of this midpoint and the Nernst peak with temperature and magnetic field. d-j) The Nernst signal $N$ and $\rho_{xx}$ vs. temperature at different magnetic fields, both the Nernst peak and the resistive transition vanish at $B=0.1$ T.}
\label{fig:1}
\end{figure*}
Fig. \ref{fig:1} presents our data on two-dimensional Nb-doped strontium titanate (STO). The heterostructure consisted of 1$\%$ at. Nb:SrTiO$_3$ ($n_{2D}$= 8.6$\times 10^{13}$cm$^{-2}$) with a thickness of 4.5 nm sandwiched by cap and buffer undoped STO layers (see Fig. \ref{fig:1}a). Previous studies documented the normal-state~\cite{Kozuka2009,Kim2011} and the superconducting properties~\cite{Kim2012} of such $\delta$-doped samples in detail. Using a standard two-thermometers-one-heater set-up (see Fig. 1b), we measured diagonal (resistivity and thermopower) as well as off-diagonal (Nernst and Hall effects) transport coefficients of the sample with the same electrodes (see the supplement~\cite{SM} for more details). As seen in panels d-j of the same figure, a Nernst signal emerges in the vortex state and its peak shifts with magnetic field and remains close to the midpoint of the resistive transition.
\begin{figure}
\centering
\includegraphics[width=0.45\textwidth]{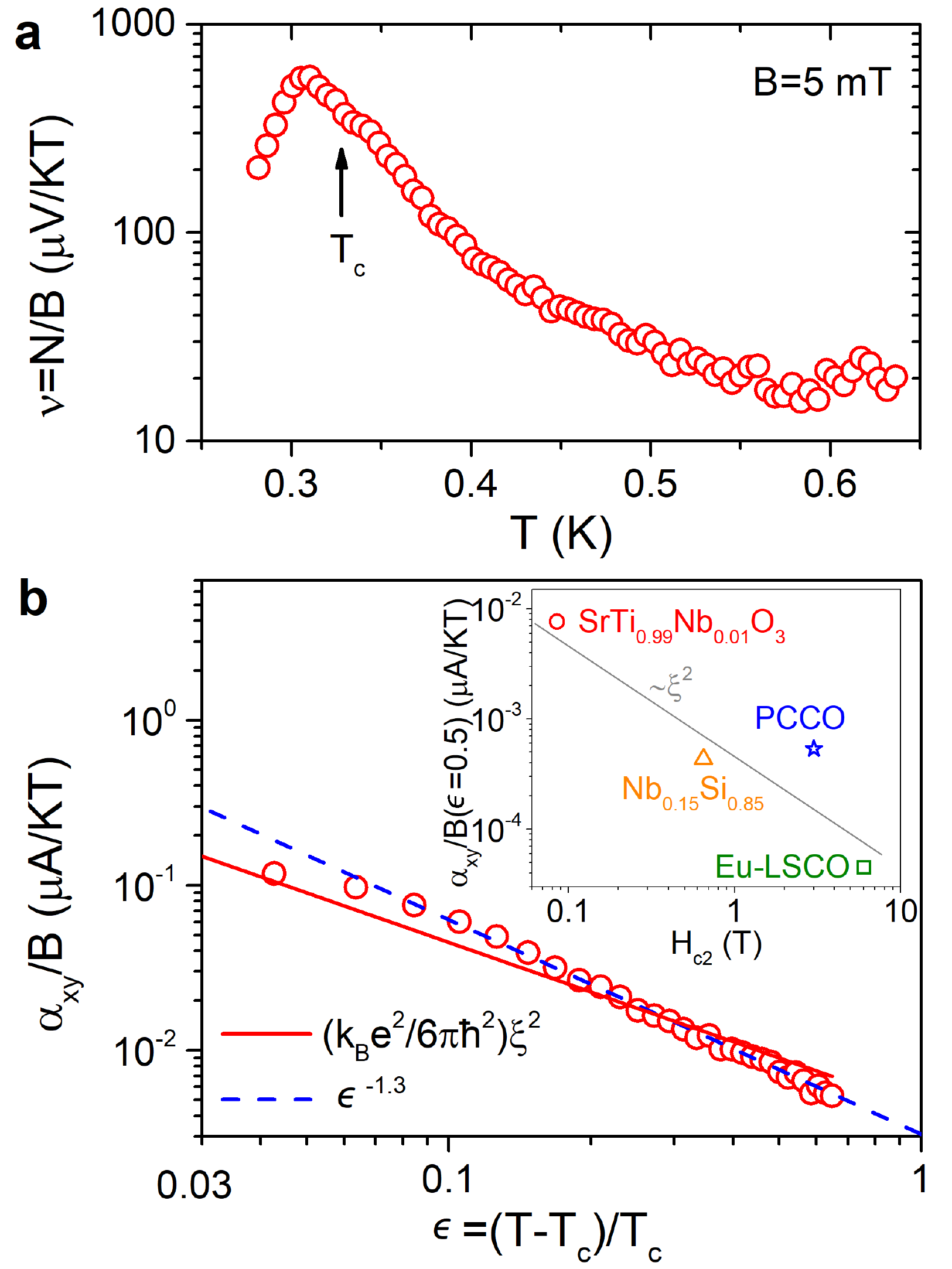}
\caption{\textbf{Nernst response in the normal state due to superconducting fluctuations:} a) The Nernst coefficient as a function of temperature in two-dimensional Nb:STO across the critical temperature at $B=5$ mT.  b) The off-diagonal component of the thermoelectric tensor, $\alpha_{xy}=\frac{N }{\rho}a_0$ as a function of reduced temperature, $\epsilon=(T-T_c)/T_c$. The dashed line represents $\epsilon^{-1.3}$, the solid line what is expected by Eq. \ref{Eq:alpha}. The inset compares the magnitude of normal-state $\alpha_{xy}$ (at $T=1.5 T_c$) in different superconductors~\cite{Pourret2006,Chang2012,Tafti2014} as a function of their upper critical field, H$_{c2}$. The dashed line represents the magnitude expected by Eq. \ref{Eq:alpha} and a coherence length given by $H_{c2}(0)$.}
\label{fig:Fig-fluc} 
\end{figure}

Fig. \ref{fig:Fig-fluc}a shows the evolution of the low-field Nernst coefficient ($\nu=N/B$) across $T_c$. Its magnitude is extremely sensitive to magnetic field. The Nernst coefficient of the normal quasi-particles detected in bulk crystals of doped STO~\cite{Lin2013} is much smaller and has an opposite sign ($\nu=-0.04$ $\mu$V/KT at $B=1$ T and $T=0.5$ K)~\cite{Lin2013,Behnia2016}. It is negligible at $B=0.005$ T. As indicated by a recent study on NbSe$_2$~\cite{XinQiLi2020}, confinement to two dimensions facilitates the observation of the superconducting contribution to the Nernst response.

Theoretically, the Nernst signal due to the Gaussian fluctuations of the superconducting order parameter  ~\cite{Ussishkin,Serbyn2009,Michaeli2009} leads to a simple expression for  the off-diagonal component of the thermoelectric tensor, $\alpha_{xy}$: 
\begin{equation}
\frac{\alpha_{xy}^{Fl}}{B} (T)=\frac{k_B e^2}{6 \pi \hbar^2} \, \xi^2(T).
\label{Eq:alpha}
\end{equation}

Here, $\xi(T)= \xi_0/\sqrt\epsilon$ is the  superconducting coherence length and   $\epsilon=(T-T_c)/T_c$ is the reduced temperature. Combining our Nernst and resistivity data, we can plot $\alpha_{xy}$ in Fig. \ref{fig:Fig-fluc}b. Its magnitude  at twice $T_c$ is compatible with what is expected  by Eq. \ref{Eq:alpha} and the zero-temperature coherence length extracted from the upper critical field ($\xi_0= 60$ nm)~\cite{Kozuka2009}. Similar observations were previously reported for amorphous superconductors~\cite{Pourret2006,Pourret2007,Spathis2008} and in cuprates~~\cite{Chang2012,Tafti2014}. Because of the long $\xi$, $\alpha_{xy}$ found here is larger than those studied previously (See the inset in Fig. \ref{fig:Fig-fluc}b and the supplement~\cite{SM}). 

\begin{figure}
\includegraphics[width=\linewidth]{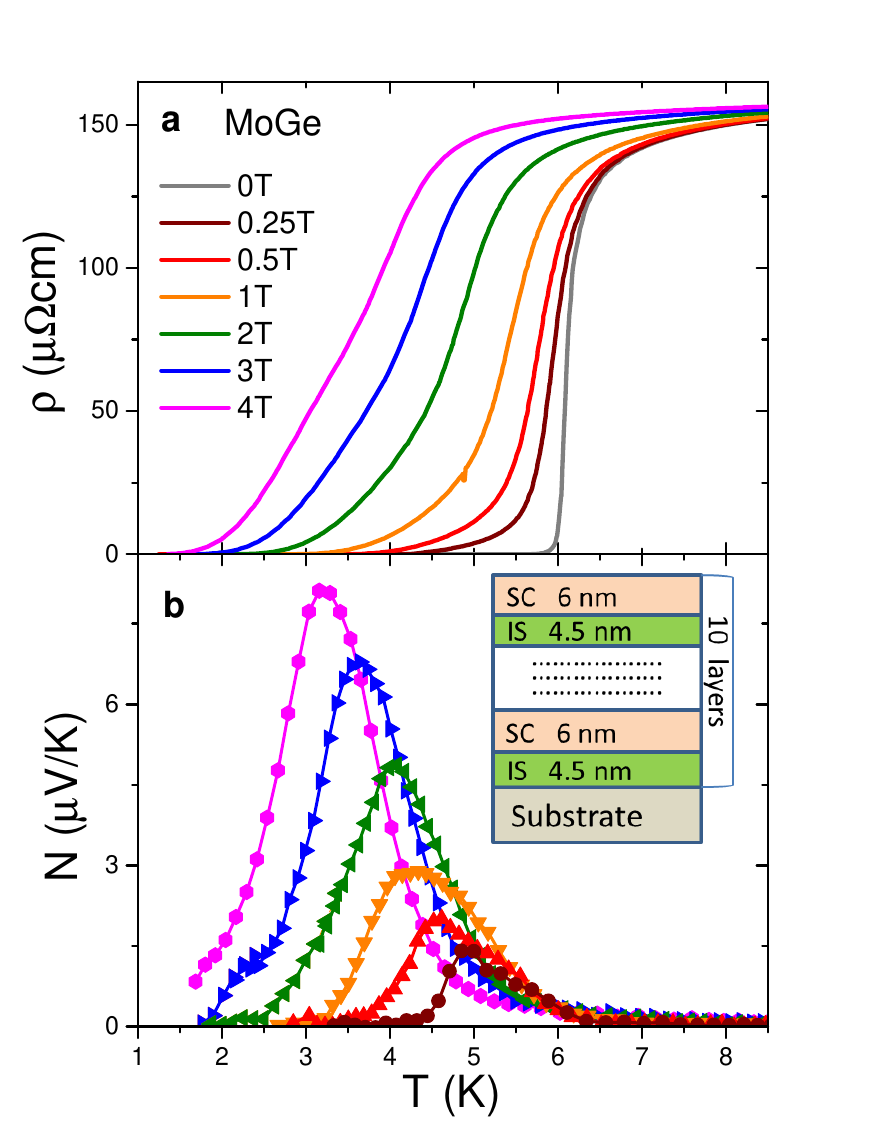}
\caption{\textbf{Nernst effect in amorphous MoGe:} a) Evolution  of the resistive superconducting transition in amorphous films of MoGe with magnetic field. b) Nernst effect $N$ in the same sample. The color code for magnetic fields is identical to the one used in the upper panel. The inset schematically depicts the structure of the sample consisting of alternating  superconducting and insulating layers.}
\label{fig:2}
\end{figure}

\begin{figure*}
\centering
\includegraphics[width=\linewidth]{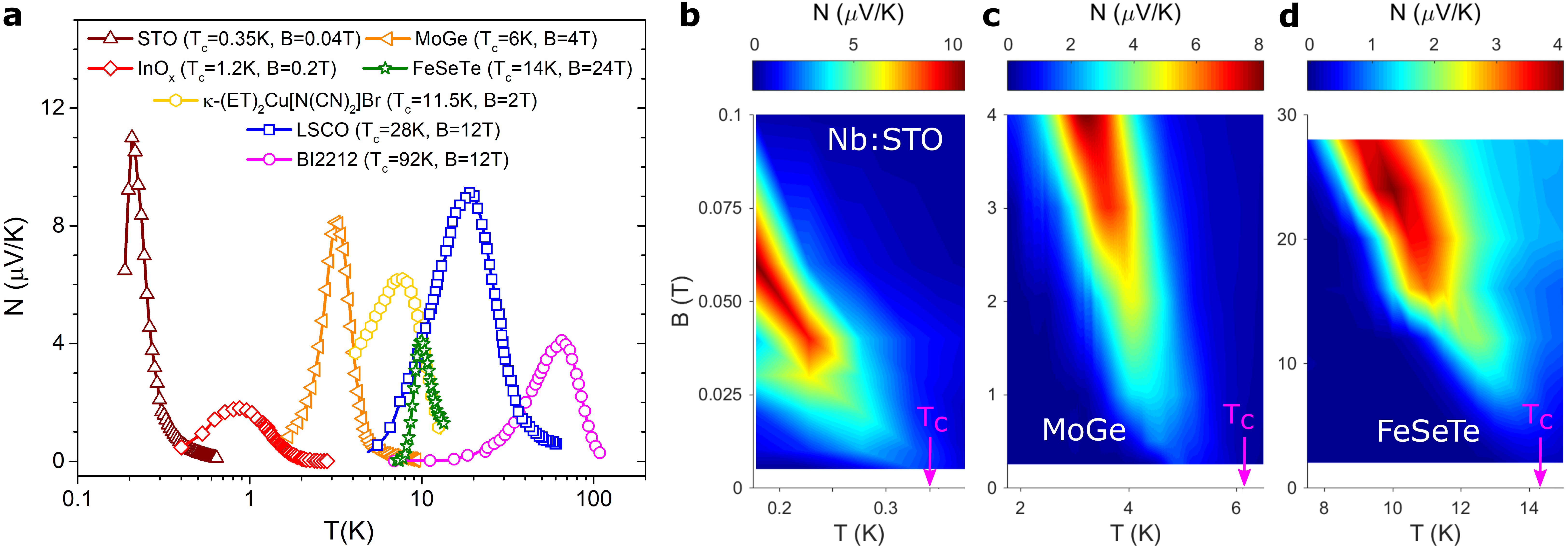}
\caption{\textbf{Peak Nernst signal in different superconductors:} a) The Nernst signal as a function of temperature in STO and MoGe (present work) compared with data on an amorphous film of InO$_x$~\cite{Spathis2008}, on FeSe$_{0.6}$Te$_{0.4}$~\cite{Pourret2011}, on $\kappa$-(ET)$_2$Cu[N(CN)$_2$]Br~\cite{Logvenov1997}, on La$_{1.92}$Sr$_{0.08}$CuO$_4$~\cite{Capan2002} and on Bi$_2$Sr$_2$CaCu$_2$O$_{8+\delta}$~\cite{Li1994}. For each system, the critical temperature is indicated together with the magnetic field at which the observed peak was the largest. In all systems, this magnetic field is the one at which the peak is largest, except in Bi2212~\cite{Li1994}, for which the data was restricted to 12 T. The field and temperature dependence of the Nernst signal are shown as color plots for b) Nb:STO, c) MoGe and d) FeSe$_{0.6}$Te$_{0.4}$~\cite{Pourret2011}. Note the similarity in the peak amplitude in contrast to the large difference in the field and temperature scales.}
\label{fig:3}
\end{figure*}

We now turn our attention to the vortex Nernst signal below the critical temperature. The Nernst signal in Nb:STO peaks to $\approx 11\mu$V/K (see Fig. 1h,i and Fig. 4b). At the temperature and magnetic field of this peak, the measured resistivity is $\approx100 \mu\Omega$cm. Therefore, the peak transverse thermoelectric response is $\alpha_{xy}=N/\rho=11$ A/Km. In the traditional approach to the vortex dynamics~\cite{Huebener1979,Li1994,Behnia2016}, this is set by a balance between the thermal force (proportional to the entropy of each vortex, S$_d$) and the Lorentz force proportional to its magnetic flux, $\phi_0=h/2e=2.07\times 10^{-15}$ Tm$^2$~\cite{Tinkham1996}. This yields S$_d = \phi_0\times \alpha_{xy} \approx 2.3 \times 10^{-14}$ J/K. m~\cite{SM}.

Sergeev and co-workers~\cite{Sergeev_2010}, after commenting the inadequacies of previous theories~\cite{Stephen1966,Maki1971},  (See the supplement for details) proposed the following expression for vortex transport entropy:  
\begin{equation}
S^{core}_{d} \simeq -\pi \xi^2 \frac{\partial }{\partial T}\frac{H_c^2}{8\pi}
\label{serg}
\end{equation}

The right side of Eq.\ref{serg} is the product of the vortex size ($\xi$ is the coherence length) and the  entropy difference between the two competing phases. Indeed, the thermodynamic critical field, H$_c$, is set by the difference between the free energies of the normal, F$_n$ and the superconducting, F$_s$ phases set : $\frac{H_c^2}{8\pi}=F_n-F_S$~\cite{Tinkham1996}. Using the experimentally known coherence length and critical fields, Eq.\ref{serg} yields  S$_d = 1.2 \times 10^{-12}$ J/K. m (see the supplement for details~\cite{SM}),  fifty times larger than the experimental value and indicating the absence of a crucial ingredient. 

Fig. \ref{fig:2} presents a study of the Nernst effect in another two-dimensional superconductor, namely amorphous MoGe, a platform for studying superconductor-insulator transitions~\cite{Yazdani1995}. The Nernst peak evolves concomitantly with the resistive transition with increasing magnetic field. The vortex Nernst signal peak is slightly lower than the peak in Nb: STO. The extracted $\alpha_{xy}$ ($\approx$ 14 A/Km) and S$_d (\approx$ 2.8 $\times 10^{-14}$ J/K.m) are almost the same. In other words, these two superconductors, despite an almost 20-fold difference in their  $T_c$s (6.2 K \textit{vs.} 0.34 K) and their H$_{c2}$s  (7 T \textit{vs.} 0.1 T) have similar entropy per vortex.

Fig.\ref{fig:3}a) shows the Nernst  data in a number of superconductors. Some are layered, others isotropic. Some are crystalline, others amorphous. Some are conventional, others unconventional. Some were studied as thin films, others as single crystals. In spite of the large difference in the critical temperature, the Nernst signal in all peaks to a few $\mu V/ K$.  Figs. \ref{fig:3}b-d compares the contours of $N(T, B)$ in three different superconductors. The field and the temperature scales differ by two orders of magnitude, but the summit has a comparable magnitude of 4 -10 $\mu$V/K. This similarity in the magnitude of the vortex Nernst response below $T_c$ is to be contrasted with the material-dependent amplitude of the fluctuating Nernst signal above $T_c$ and the material-dependent amplitude of the quasi-particle Nernst signal. The latter is known to spread over six orders of magnitude in different metals~\cite{Behnia2009,Behnia2016}.  Theory gives a satisfactory account of the amplitude of the quasi-particle or the fluctuating Nernst signal but, as we saw above, not the vortex Nernst signal.

One defect of the common picture of the vortex Nernst signal is its neglect of forces other than the thermal force acting on a vortex as discussed in the supplement \cite{SM}. An upper boundary to $N$ is equivalent to a lower boundary to the viscosity-to-entropy density ratio for the vortex liquid. Such a boundary is a subject of current interest~\cite{Kovtun2005,Trachenko2020} also discussed in the supplement~\cite{SM}. 

\begin{table}
 \caption{The peak Nernst signal in superconductors belonging to four different families: SrTi$_{0.99}$Nb$_{0.01}$O$_3$ (Nb:STO), FeSe$_{0.6}$Te$_{0.4}$(FeSeTe)~\cite{Pourret2011}, $\kappa$-(ET)$_2$Cu[N(CN)$_2$]Br($\kappa$-ET)~\cite{Logvenov1997} and La$_{1.92}$Sr$_{0.08}$CuO$_4$(LSCO08)~\cite{Capan2002}. Also listed are sheet resistance per layer (resistivity divided by the lattice parameter along the orientation of magnetic field) measured at the temperature and the magnetic field corresponding to N=N$^{peak}$ and the deduced entropy per vortex per layer (see the supplement~\cite{SM} for a discussion of the available Nernst data).}
 \begin{ruledtabular}
\begin{tabular}{lcccccc}
Compound & $T_{c}$ & N$^{peak}$& c& $\rho^{peak}$/c & S$_{d}^{sheet}$ \\
            & [K] & [$\mu$/K] & nm & [k$\Omega$]  & [10$^{-23}$J/K]  \\
\hline
Nb:STO  & 0.35 & 11 &  0.39 &2.6 & 0.89  \\
FeSeTe & 14 & 4  &  0.58 & 0.86 & 0.96  \\
$\kappa$-(ET) & 11 &6.1 &2.9 &  1.31 & 0.96  \\
LSCO08  & 29 & 9.1 & 1.2&2.12 & 0.88  \\
\hline
\end{tabular}
\end{ruledtabular}
\label{Tab1}
\end{table}

Table \ref{Tab1} lists four different crystalline superconductors and their largest value of the Nernst signal at any field and temperature, $N^{peak}$. They belong each to a different family and they are chosen because the resistivity of the sample at $N= N^{peak}$ (dubbed $\rho^{peak}$) has been reported, allowing to calculate the vortex entropy per layer, using the lattice parameter $c$: S$_{d}^{sheet}=\Phi_0\frac{N^{peak}}{\rho^{peak}}c$. As seen in the table, S$_{d}^{sheet}$ is similar and of the order of k$_B ln2= 0.95 \times 10^ {-23}J/K$. In other words, despite the dissimilarity in the coherence length and in the penetration depth, the entropy carried by each vortex per sheet is of the order of a Boltzmann constant. 

Our observation implies that Eq.\ref{serg} does not give an accurate account of the mobile entropy of a superconducting vortex and the problem should be deeply reconsidered. At this stage, we can identify two obvious shortcomings with equation. First it  assumes that the entropy density in the vortex core is identical to the entropy density in the normal phase. This neglects the existence of the  Caroli–de Gennes–Matricon~\cite{Caroli1964} levels in the core. Second, it takes for granted that all core entropy is mobile and does not distinguish between what is bound to a mobile flux line and what is not.

To sum up, we find that in four superconductors with different normal states, pairing symmetries and critical temperatures, the Nernst transport entropy per vortex per layer is of the order of k$_B$. We expect this to motivate experimental studies of the vortex Nernst signal in other superconductors of interest~\cite{He2013,Drozdov2015,Cao2018}.  

We thank H. Aubin, M. V. Feigel'man, S. A. Hartnoll, S. A. Kivelson K. Trachenko A. A. Varlamov and G. E. Volovik for discussions.  CWR acknowledges the support of Fonds-ESPCI, Paris. This work was supported by a ‘QuantEmX’ Exchange Awards at Stanford University (KB) and at ESPCI (AK), by the Agence Nationale de la Recherche  (ANR-18-CE92-0020-01; ANR-19-CE30-0014-04) and by Jeunes Equipes de l$'$Institut de Physique du Coll\`ege de France. HI, MK, CB, and HYH were supported by the U.S. Department of Energy, Office of Basic Energy Sciences, Division of Materials Sciences and Engineering, under contract no. DE-AC02-76SF00515. AK was supported by the National Science Foundation Grant NSF-DMR-1808385.

\bibliography{biblio}

\newpage
\appendix
\section{Experimental technique}
The two-thermometers-one-heater setup shown in Fig. 1b of the main text permits the measurement of all transverse and longitudinal electric and thermoelectric transport coefficients in the same conditions. A longitudinal thermal gradient $\nabla_{x}T$ is generated by gluing one end of the sample with silver paste to a cold finger and connecting the other end to a heater. The thermal gradient $\nabla_{x}T=(T_{\textnormal{hot}}-T_{\textnormal{cold}})/s$ was measured by two RuO$_2$ thermometers $T_{\textnormal{hot}}$ and $T_{\textnormal{cold}}$ attached to the sample with electric leads, separated by a distance $s$, that allowed as well to measure the longitudinal and transverse voltage drops, $V_{x}$ and $V_{y}$, respectively. The Nernst coefficient is obtained using $N=E_y/\nabla_{x}T$ with $E_y=V_y/w$ and the sample width $w$. By applying an electric current instead of a heat current, the same experimental setup also allows to measure the longitudinal resistance and the Hall effect. The magnetic field was applied perpendicular to the orientation of both the applied heat current and the measured voltage drops.
The setup was mounted on a $^4He-^{3}$He-dilution probe as well as on a home-built measurement stick for a Physical Property Measurement System (PPMS) that allowed to access temperatures down to 1.7 K.

\begin{figure}
\includegraphics[width=0.5\textwidth]{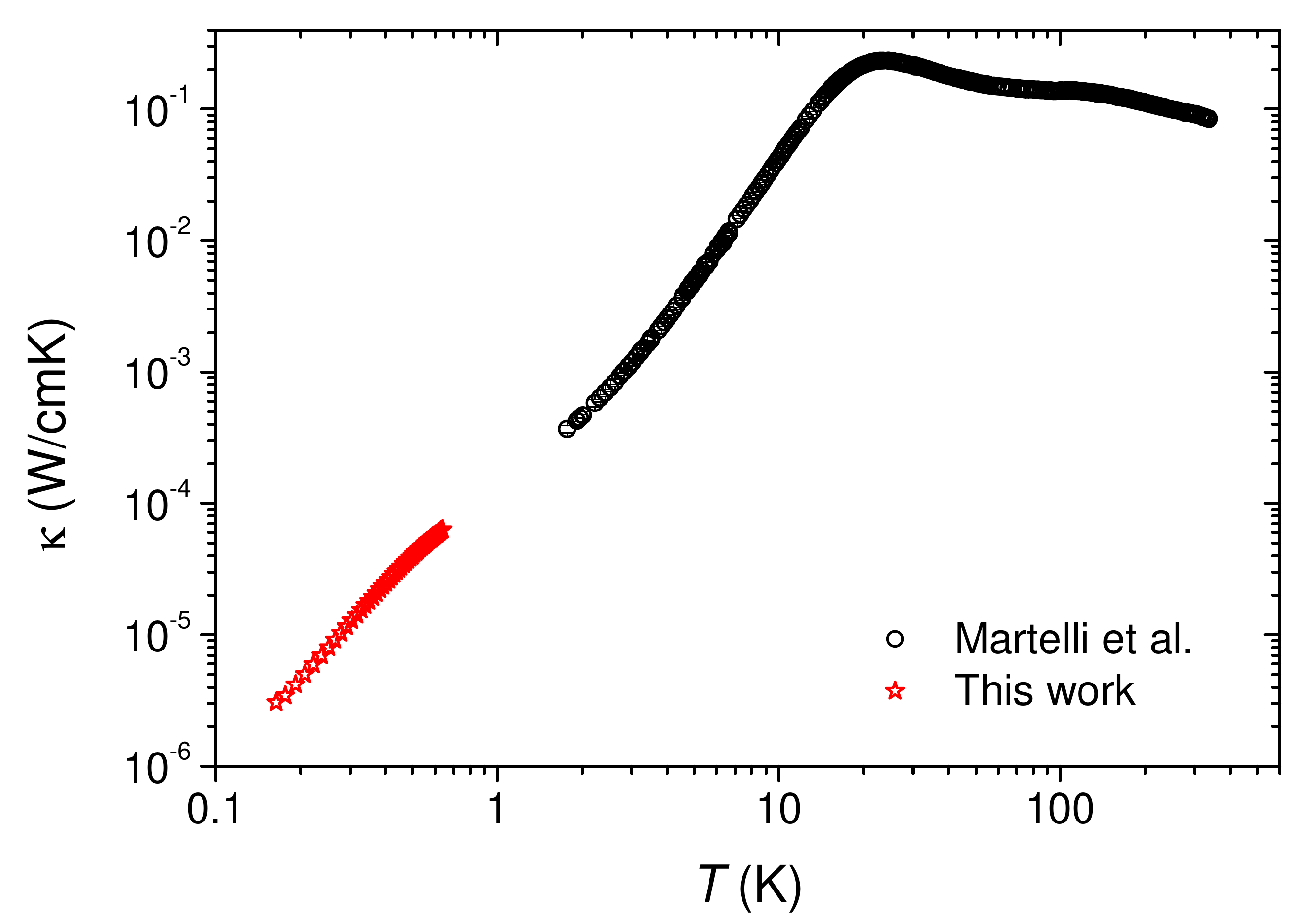}
\caption{\textbf{Thermal conductivity} Thermal conductivity $\kappa$ as a function of temperature $T$ in comparison to previous data on SrTiO$_3$ by Martelli et al. \cite{Martelli2018}.}
\label{fig:SI1}
\end{figure}

Prior to the Nernst measurements, we determined the thermal conductivity $\kappa$ of the insulating substrate by measuring the temperature gradient introduced by a measured heat current. Fig. S \ref{fig:SI1} compares the measured $\kappa$ of the substrate  (i.e., an insulating SrTiO$_3$ crystal) with previous measurements of bulk SrTiO$_3$ above 1.8 K~\cite{Martelli2018}. The two data sets appear to join each other smoothly. This agreement confirms the accuracy of our quantification  of the thermal gradient.

The presence of a sub-micronic thick $\delta$-doped  SrTi$_{0.99}$Nb$_{0.01}$O$_3$ sample on the substrate does not alter heat transport in any detectable way. Knowing the thermal conductivity, which remains unchanged by the application of magnetic fields smaller than 0.1 T, allowed us to quantify the Nernst signal by measuring the transverse electric field $E_y$ produced by a specific longitudinal thermal current at a specific temperature.

\section{The Nernst signal generated by short-lived Cooper pairs}

Fluctuations of the superconducting order parameter above the critical temperature generate a Nernst signal. This was first theoretically described by Ussishkin, Sondhi and Huse (USH)~\cite{Ussishkin} and was then elaborated in more detail and extended to finite magnetic fields by Serbyn and co-workers~\cite{Serbyn2009} and by Michaeli and Finkel$'$stein~\cite{Michaeli2009}. Its experimental relevance was tested in both amorphous superconductors~\cite{Pourret2006,Pourret2007,Pourret2009,Behnia2016} and in high-$T_c$ cuprates~\cite{Chang2012,Tafti2014,Cyr2018}. In this section, we compare the magnitude of the Nernst signal above $T_c$ in the Nb-doped STO film of the present study with previous reports.\\
\begin{figure}
\includegraphics[width=0.5\textwidth]{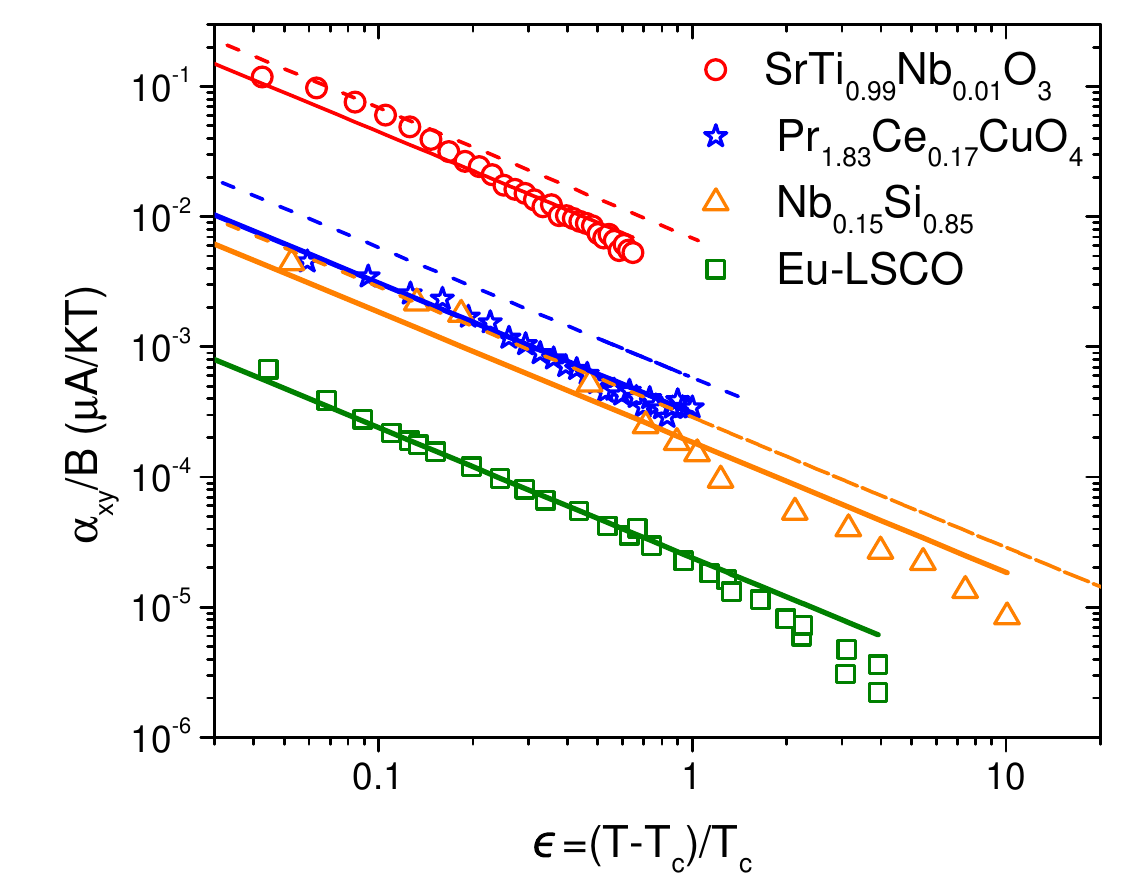}
\caption{\textbf{Fluctuating Nernst signal in different superconductors:} The transverse thermoelectric conductivity divided by magnetic field ($\alpha_{xy}/B$) for the Nb-doped STO hetero-structure,  Nb$_{0.15}$Si$_{0.85}$ \cite{Pourret2006},  Pr$_{1.83}$Ce$_{0.17}$Cu$_{4}$O (PCCO) \cite{Tafti2014} and La$_{1.69}$Eu$_{0.2}$Sr$_{0.11}$CuO$_{4}$ (Eu-LSCO) \cite{Chang2012}. Solid lines show linear fits to the data as expected by Eqs. \ref{Eq:SI3} and \ref{Eq:SI4}.} 
\label{fig:SI2}
\end{figure}
The USH expression for superconducting fluctuations for a 2D superconductor is remarkably simple. The off-diagonal component of the thermoelectric tensor is expected to have an additional contribution which scales with the superconducting coherence length $\xi$. Provided that the magnetic field is small enough (i.e., $B<<\phi_{0}/2\pi\xi^2$), one expects:

\begin{equation}
\frac{\alpha_{xy}^{Fl}}{B}=\frac{k_B e^2}{6 \pi \hbar^2} \, \xi^2
\label{Eq:SI3}
\end{equation}

The Ginzburg-Landau superconducting correlation length in the normal state~\cite{Larkin2005} is:

\begin{equation}
\xi=\frac{\xi_0}{\sqrt{\epsilon}}
\label{Eq:SI4}
\end{equation}

Here, $\xi_0$ is the temperature-independent superconducting coherence length and  $\epsilon=(T-T_c)/T_c$ is the reduced temperature. Thus, the USH expression expects a fluctuating signal  proportional to $1/\epsilon$.

Fig. S \ref{fig:SI2} compares the experimental $\alpha_{xy}/B (\epsilon)$ in Nb-doped STO  with three other superconductors, namely Nb$_{0.15}$Si$_{0.85}$ \cite{Pourret2006}, Pr$_{1.83}$Ce$_{0.17}$Cu$_{4}$O (PCCO) \cite{Tafti2014} and La$_{1.69}$Eu$_{0.2}$Sr$_{0.11}$CuO$_{4}$ (Eu-LSCO) \cite{Chang2012}.  In all superconductors, the data roughly follows a $1/\epsilon$-dependence confirming USH theory. 

The deviation from this  behavior at low $\epsilon$ may result from short wavelength  effects around  $\epsilon =0.25$~\cite{Cimberle1997}.

According to Eq. \ref{Eq:SI4}, $\alpha_{xy}^{Fl} (\epsilon=1$) should be larger in a superconductor with longer $\xi_0$. As seen in Fig.  S\ref{fig:SI2}, this is indeed the case. The signal  in Nb-doped STO is two orders of magnitude larger than in Eu-LSCO. The latter has an upper critical field which is almost two orders of magnitude larger.

It is instructive to compare  the superconducting coherence lengths extracted from three distinct experimental sources:

i) The  expression for the coherence length in a dirty superconductor, $\xi_{0d}$, extracted from Fermi velocity $v_F$ and  mean-free path $\ell$, \cite{Larkin2005}: 
\begin{equation}
\xi_{0d}= 0.36\sqrt{\frac{3\hbar v_F \ell}{2k_B T_c}}
\label{Eq:SI5}
\end{equation}

ii)  The zero-temperature upper critical field: 
\begin{equation}
\xi_0 = \sqrt {\frac{\phi_0}{2 \pi H_{c2}(0)}}
\label{Eq:SI6}
\end{equation}

iii) The Nernst data using  equations \ref{Eq:SI3} and \ref{Eq:SI4}:
\begin{equation}
\xi_N = \sqrt {\frac{6 \pi \hbar^2}{k_B e^2} \frac{\alpha_{xy}^{Fl}(\epsilon=1)}{B}}
\label{Eq:SI7}
\end{equation}
Table \ref{TabSI1} presents such a comparison. As seen in the table, the three numbers are close but not identical.  For Nb-doped STO, $v_{F}l$ was estimated via $v_{F}l=3\kappa/\gamma_e T=(\pi k_B/e)^2\sigma/\gamma_e$ with the measured conductivity $\sigma=1/\rho_{xx}$ and the electronic specific heat taken from ref. \cite{Lin2014}.

 \begin{table*}
 \centering
 \caption{\textbf{A comparison of four superconductors:}  Physical properties of superconductors in which a fluctuating Nernst signal above $T_c$ has been detected. The superconducting coherence length has been extracted using three distinct equations Eq. \ref{Eq:SI5}, Eq. \ref{Eq:SI6} and Eq. \ref{Eq:SI7}. }
 \begin{ruledtabular}
\begin{tabular}{lcccccc}
Compound & $T_{c}$ & $H_{c2}$& $v_{F}l $ &  $\xi_{0}$& $\xi_{0d}$ &  $\xi_{N}$ \\
            & [K] & [T] & [m$^2$/s] & [nm]   & [nm]  & [nm] \\
\hline
SrTi$_{0.99}$Nb$_{0.01}$O$_3$ (present study)  & 0.32 & 0.085 & $9.3\times10^{-4} $ & 62 & 64 & 52  \\
Nb$_{0.15}$Si$_{0.85}$ \cite{Pourret2007} & 0.38 & 1.1 & $4.35\times10^{-5}$& 17  & 13 & 10  \\
Pr$_{1.83}$Ce$_{0.17}$Cu$_{4}$O (PCCO)  \cite{Tafti2014} & 19.5 & 3 & $4.5\times10^{-3}$ &  10  & 18 & 14  \\
La$_{1.69}$Eu$_{0.2}$Sr$_{0.11}$CuO$_{4}$ (Eu-LSCO) \cite{Chang2012} & 3.86 & 6  & - & 7 & - & 3.8 \\
\hline
\end{tabular}
\end{ruledtabular}
\label{TabSI1}
\end{table*}

\section{Nernst effect, resistivity and the vortex transport entropy}
Here we briefly sketch the widely used picture of the vortex Nernst response. A thermal gradient generates a thermal force on each vortex. This force is balanced by a damping force leading to a steady displacement of vortices. The finite vortex velocity will generate a phase slip along the perpendicular orientation and, thanks to the Josephson equation, a finite electric field. Now, the thermal force on a vortex with an entropy of S$_{d}$ is countered by a damping force proportional to the velocity of the vortex line $v_{L}$ and a viscous parameter $\eta'$~\cite{Huebener1979}: 

\begin{equation}
 S_{d} \overrightarrow{\nabla_{x}}T = \eta' \overrightarrow{ v}_{L} 
\label{ft}
\end{equation}

A steady motion of vortices would generate a temporal variation of the superconducting phase along the orientation perpendicular to the vortex movement. With each vortex carrying a magnetic flux of $\phi_0= \frac{h}{2e}$ and their density equal to $n_V$, the Josephson equation leads to :
\begin{equation}
 \overrightarrow{E}=  \overrightarrow{ v}_{L} \times  n_V \Phi_0\hat{z}
\label{Josephson}
\end{equation}

By combining the two equations and introducing the entropy density $s=\frac{S_d}{n_V}$, one finds that the Nernst signal is  set by the ratio of entropy density to the viscous parameter $\eta'$. 

\begin{equation}
N\equiv\frac{E_y}{\nabla_x T}= \Phi_0\frac{s}{\eta'}
\label{N-eta-s}
\end{equation}

Assuming  that $\eta'$ is also the damping parameter opposing the Lorentz force in the flux flow resistivity, one van write: 
\begin{equation}
\overrightarrow{J} \times \Phi_0 \hat{z}=\eta' \overrightarrow{ v}_{L}
\label{ff1}
\end{equation}
Hosephon equation implies: 
\begin{equation}
\overrightarrow{E} = \overrightarrow{v_L}\times \overrightarrow{B}
\label{ff2}
\end{equation}

Therefore, in the flux flow regime:

\begin{equation}
\rho=\Phi_0^2\frac{n_V}{\eta'}
\label{fFR}
\end{equation}
Combining Eq.\ref{fFR} and Eq.\ref{N-eta-s}, one finds
\begin{equation}
\frac{N}{\rho}=\frac{S_d}{\phi_0}
\label{S_d}
\end{equation}

Thus, the vortex transport entropy S$_d$ can be deduced from N and $\rho$ measured at the same temperature and magnetic field.

\section{Theoretical magnitude of the  Vortex transport entropy and comparison with experiment}
Sergeev and co-workers~\cite{Sergeev_2010}, after revisiting earlier theories ~\cite{Stephen1966,Maki1971}, contested the validity of the following expression for the vortex transport entropy derived by Stephen~\cite{Stephen1966}:

\begin{equation}
S^{EM}_{d} = -\frac{\phi_0}{4\pi}\frac{\partial H_{c1}}{\partial T}
\label{s1}
\end{equation}

Such an expression is derived by assuming that the entropy of the vortex is set by the temperature derivative of the energy cost to introduce a vortex at the lower critical field, $H_{c1}$. According to ref.~\cite{Sergeev_2010}, this energy cost includes supercurrents, which do not transport entropy. Therefore, the correct expression for entropy is: 
\begin{equation}
S^{core}_{d} \simeq -\pi \xi^2 \frac{\partial }{\partial T}\frac{H_c^2}{8\pi}
\label{s2}
\end{equation}

Here, $H_{c}$ is the thermodynamic critical field  and $\xi$ is the coherence length. $S^{core}_{d}$ excludes the supercurrent contribution and is therefore smaller than $S^{EM}_{d}$ by a factor of 2$\ln \kappa$. Here,  $\kappa = (\frac{\lambda}{\xi})$ is the Ginzburg-Landau parameter and $\lambda$ is the penetration depth.  Eq. \ref{s2} can be simplified using: $H_{c1}= H_c\kappa \sqrt 2$ and $H_{c2}= \frac{ln \kappa}{\kappa \sqrt 2}$~\cite{Tinkham1996}:
\begin{equation}
S^{core}_{d} \simeq \frac{1} {2ln\kappa}\frac{\Phi_0}{4 \pi}\frac{\partial H_{c1} }{\partial T}
\label{s2b}
\end{equation}

The lower critical field of bulk superconducting STO has been the subject of a detailed study~\cite{Collignon2017}. For  1$\%$  Nb doping level ($n_{3D}=1.9  \times 10^{20}$ cm$^{-3}$), it was found to be $H_{c1}$(0)= 4.8 Oe. The $H_{c1}(T)$ curve allows to extract the slope ($\frac{\partial H_{c1}}{\partial T}(T=0.2 k)= -25$ Oe/K). Inserting this in Eq. \ref{s1}, one finds : 
\begin{equation}
S^{EM}_{d}=  5.2 \times 10^{-12}\,\textnormal{J/K. m}\
\end{equation}

The Ginzburg-Landau parameter of STO at this carrier concentration was estimated  to be $\frac{\lambda}{\xi} \simeq 8.5$~\cite{Collignon2017}. Thus, the core entropy $S^{core}_{d}$ would be ${2 \ln\kappa}\simeq 4.3$ times smaller  :
\begin{equation}
 S^{core}_{d} \approx  1.2 \times 10^{-12}\,\textnormal{J/K. m}\
\end{equation}
This is close to a generic estimation by Sergeev \textit{et al.}~\cite{Sergeev_2010} (S$^{EM}_{d} = 1.6 \times 10^{-7}$ erg/K. cm). 

However, as discussed in the main text, the experimental value combining the measured amplitudes of the Nernst signal and resistivity ($S_{d} (exp.) \approx  2.2 \times 10^{-14}$J/K. m) is fifty times lower.

\section{Other forces acting on vortices}

The treatment of the vortex Nernst response may need the inclusion of all forces on a vortex besides the thermal force acting on the core entropy. Momentum exchange occurs between three subsystems: the superfluid `vacuum'~\cite{Volovik2003}, the normal quasi-particles and the topological texture introduced by the presence of vortices. The balance of all dissipative and reactive forces on a vortex~\cite{Kopnin1995,Sonin1997,Volovik2003} would lead to :

\begin{equation}
\hat{z} \times  (\overrightarrow{v}_L-\overrightarrow{v}_{s})+ d_{\perp}\hat{z} \times  (\overrightarrow{v}_n-\overrightarrow{v}_{L})+ d_{\parallel} (\overrightarrow{v}_n-\overrightarrow{v}_{L})=0
\label{Vol}
\end{equation}

This expression includes the Magnus force between the vortex and the superfluid (which has a velocity of $v_s$), the Iordanskii force (proportional to  the differential velocity of the normal fluid $v_n$ and the superfluid) and the Kopnin force (proportional to the differential velocity of the normal fluid and the vortex), which is a consequence of spectral flow~\cite{Caroli1964,Kopnin1995} in fermionic superfluids. The two dimensionless parameters $d_{\perp}$ and $d_{\parallel}$ represent dissipation and quantify the viscous response. Now, the Lorentz force generated by a charge current and a thermal force generated by a thermal gradient do not affect the three velocities in the same way. Therefore, the dissipation they cause and the associated $\eta'$ are not necessarily identical.

\section{The viscosity-entropy density ratio bound}

According to available experimental data, the vortex Nernst peak remains bellow 11 $\mu$V/K$\approx$ k$_B/8e$. This combined  with Eq. \ref{N-eta-s}, leads us to:
\begin{equation}
\frac{\eta'}{s} > c_{vo} \frac{\hbar}{k_B} 
\label{etas}
\end{equation}

Empirically, $c_{vo}\approx 8 \pi$. The viscous parameter, $\eta'$, first introduced by Bardeen and Stephen to quantify flux flow resistivity~\cite{Bardeen1965} is to be distinguished from the dynamic viscosity, $\eta$ of a compressible fluid. While $\eta'$ is the ratio of the force to velocity, $\eta$ is ratio of the force to the gradient of velocity. 

A bound to the viscosity-entropy ratio was proposed in the context of strongly interacting quantum field theories~\cite{Kovtun2005} and exists in common liquids~\cite{Trachenko2020}.  A recent theoretical work~\cite{Trachenko2020} has argued that this minimum arises because of a universal bound to the kinematic viscosity of liquids.

\section{Data on N$^{peak}$ in different families of superconductors }

To the best of our knowledge in no superconductor hitherto explored, the vortex Nernst signal exceeds $\approx 10 \mu V/ K$. We did not find any published record claiming this statement. 

A second and a stronger statement concerns a subset of studies in which: a) the maximum vortex Nernst signal as a function of field and temperature has been clearly identified; and b) the electrical resistivity at this field and temperature has been measured. These  cases are presented in table 1 of the main text.  In these cases the combination of the Nernst and resistivity data yields similar S$_{d}^{sheet}$ as seen in the table 1 of the main text. Below, we present a brief review of literature:

\textbf{Cuprates} - We have used the data by Capan \textit{et al.} \cite{Capan2002}, because it corresponds to the two criteria listed above. The largest vortex Nernst signal is N$^{peak}$= 9.1 $\mu$V/K. It occurs at $T= 9$ K and $B= 12$ T. Resistivity is reported at this temperature and magnetic field to be 280 $\mu\Omega$cm. The combination allows to extract the vortex entropy per layer. There are numerous other studies of the Nernst effect in cuprates. As early as 1990, Palstra \textit{et al.} ~\cite{Palstra1990} reported on the Ettingshausen coefficient and the thermal conductivity of YBa$_2$CuO$_3$O$_7$ in the vortex liquid state. Using their thermal conductivity data and the Bridgman relation~\cite{Bridgman,Behnia2015b}, one finds that  N$^{peak}\approx 4 \mu V$/K,  of the same order of magnitude found in the following studies directly measuring the Nernst response. The table below gives an account of what is reported by studies directly measuring the vortex Nernst signal. 

\begin{table}
 \caption{The Nernst signal observed in the vortex liquid state of various cuprates has a peak value, N$^{peak}$, observed at a specific magnetic field, B$^{peak}$. The table lists N$^{peak}$ and B$^{peak}$ as well as the the highest magnetic field explored in each study. $\rho^{peak}$ refers to the resistivity of the sample at the temperature and the magnetic field at which N$^{peak}$ was observed. In most studies, it was not specified. Note that in all cuprates N$^{peak}$ remains in the range of $\mu$ V/K.}
 \begin{ruledtabular}
\begin{tabular}{lcccccc}
Compound & Ref. & N$^{peak}$& B$^{peak}$ &Field range  & $\rho^{peak}$ &  \\
            &  & [$\mu$/K] & T& T  &[$\mu \Omega$ cm]  &  \\
\hline
La$_{1.92}$Sr$_{0.08}$CuO$_4$  & \cite{Capan2002}\ & 9.1 &12 &$<26$ & 280  \\
La$_{1.88}$Sr$_{0.12}$CuO$_4$  & \cite{Wang2001}\ & 6.5 &13.5 &$<14$ & ?  \\
La$_{1.8}$Sr$_{0.2}$CuO$_4$  & \cite{Wang2002}\ & 2.5 &11  &$<33$ & ? \\
YBa$_{2}$Cu$_{3}$CuO$_{7-\delta}$  & \cite{Li1994}\ & 2.6 &12  &$<12$ & 35 \\
YBa$_{2}$Cu$_{3}$CuO$_{6.6}$  & \cite{Rullier2006}\ & 4.5 &8  &$<8$ & ? \\
YBa$_{2}$Cu$_{3}$CuO$_{6.67}$  & \cite{Cyr2018}\ & 4.1 &9  &$<15$ & ? \\
YBa$_{2}$Cu$_{3}$CuO$_{6.99}$  & \cite{Wang2006}\ & 3.9 &14  &$<14$ & ? \\
Bi$_{2}$Sr$_{2}$CaCu$_{2}$O$_{8+x}$  & \cite{Li1994}\ & 4.1 &12  &$<12$ & 55 \\
Bi$_{2}$Sr$_{2}$CaCu$_{2}$O$_{8+x}$  & \cite{Wang2006}\ & 3.1 &14  &$<14$ & ? \\
Pr$_{1.85}$Ce$_{0.15}$CuO$_{4}$  & \cite{Balci2003}\ & 1.8 &1  &$<9$ & ? \\
\hline
\end{tabular}
\end{ruledtabular}
\label{TabS2}
\end{table}

Most of these Nernst studies do not report concomitant  resistivity data. The study by Ri \textit{et al.}~\cite{Li1994} is a notable exception. However, it was limited to $B<12$ T, while N is still rising. Therefore, the observed N$^{peak}$ may not be the genuine N$^{peak}$ of the system possibly occurring at a higher magnetic field. Nevertheless, S$_{d}^{sheet}$ extracted from N$^{peak}$ and $\rho^{peak}$ would yield a value of the order of k$_B$.


\textbf{Organic superconductors} - We are aware of two studies on organic superconductors~\cite{Logvenov1996,Logvenov1997,Nam2007}. Logevenov \textit{et al.}~\cite{Logvenov1996,Logvenov1997} and Nam \textit{et al.}~\cite{Nam2007} studied the two members of the $\kappa$-(BEDT-TTF)$_2$X$_2$ family and both identified  N$^{peak}$ as a summit in the $(B,T)$ plane. However, Logevenov \textit{et al.} in contrast to Nam \textit{et al.} present concomitant  resistivity data, included in table 1. 

\textbf{Iron-based superconductors} - There have been several reports of the Nernst coefficient in iron-based superconductors (for example, see~\cite{Kasahara2016}). However, to the best of our knowledge, the only one which clearly identifies N$^{peak}$ in the vortex liquid together with flux flow resistivity is ref. \cite{Pourret2011}.

\textbf{Others} - A very recent study~\cite{XinQiLi2020} quantified the Nernst peak of two-dimensional crystalline NbSe to be $\sim 5$ $\mu$V/K  (in contrast to the bulk crystals, where the vortex liquid is narrow and the normal-state Nernst response is by far dominant~\cite{Bel2003}).

\end{document}